\documentclass[%
 jmp,
 amsmath,amssymb,
twocolumn,%
]{revtex4-1}

\usepackage{graphicx}
\usepackage{dcolumn}
\usepackage{bm}

\usepackage[allcolors=blue,colorlinks=true]{hyperref}
\usepackage[utf8]{inputenc}
\usepackage[T1]{fontenc}
\usepackage{mathptmx}
\usepackage{etoolbox}
\usepackage{diagbox}
\usepackage{subfigure}
\usepackage{float}
\usepackage[dvipsnames]{xcolor}
\usepackage{makecell} 

\DeclareSymbolFont{newfont}{OML}{cmm}{m}{it}
\DeclareMathSymbol{\Epsilon}{3}{newfont}{15}

\makeatletter
\def\@email#1#2{%
 \endgroup
 \patchcmd{\titleblock@produce}
  {\frontmatter@RRAPformat}
  {\frontmatter@RRAPformat{\produce@RRAP{*#1\href{mailto:#2}{#2}}}\frontmatter@RRAPformat}
  {}{}
}%
\makeatother
\usepackage{xr}
\usepackage{graphicx} 
\begin{document}

\begin{abstract}
Simulating interactions between non-spherical colloidal particles is computationally challenging due to the complex dependency of forces and energies on their geometry.
We introduce and evaluate both descriptor-based and end-to-end models for predicting interaction energies and forces. Then, we compare various descriptors coupled with different regression models, like Behler-Parinello descriptors, Smooth Overlap of Atomic Positions, and neuroevolution potential, as well as multiple end-to-end models, namely SchNet, DimeNet, and DimeNet++.
Among these, the neuroevolution potential (NEP) offers an optimal balance between accuracy and computational efficiency. NEP, originally developed for atomistic systems, represents interactions between rigid anisotropic bodies using point clouds, which enables the representation of any arbitrary shape. 
Molecular dynamics simulations using NEP, accurately reproduced structural properties across diverse particle shapes including cubes, tetrahedra, pentagonal bipyramids, and twisted cylinders, while achieving roughly up to an order-of-magnitude speedup over other methods. Additionally, we show that the extension of the method to multi-face shapes with different interactions on their surface is straightforward. We used a twisted cylinder, which lacked any point group symmetry, to demonstrate the flexibility and accuracy of NEP. Our approach enables scalable simulations of complex colloidal systems and can potentially help to facilitate efficient  studies on shape dependent interactions and phase behavior in the future.
\end{abstract}

\title{Machine-Learning Potentials for Efficient Simulations of Anisotropic Colloids}

\author{B. Ru\c{s}en Argun}
\affiliation{Mechanical Engineering, 
                              The Grainger College of Engineering, University of Illinois,Urbana-Champaign, 61801, IL, United States}

\author{Antonia Statt}
\email{statt@illinois.edu}
\affiliation{Materials Science and Engineering, 
                              The Grainger College of Engineering, University of Illinois, Urbana-Champaign, 61801, IL, United States}

\maketitle

\section{Introduction}
Classical particle-based simulations such as molecular dynamics (MD) and Monte Carlo (MC) methods are typically performed using spherical particles. However, many systems of interest, such as colloidal self-assembly \cite{sacanna2013shaping, cai2021colloidal, henzie2012self}, structural biology ~\cite{lewandowska2021fungicidal}, drug delivery ~\cite{hatch2024anisotropic}, granular matter ~\cite{zhao2023role}, involve non-spherical particles. 
While interactions between spherical particles depend solely on the distance between their centers, interactions between non-spherical particles are significantly more complex, posing inherent challenges for traditional simulation methods. 

Various methods have been developed to simulate non-spherical particles. For example, efficient overlap detection algorithms enable the simulation of anisotropic shapes\cite{gilbert1988fast,damasceno2012predictive}, however, they are generally limited to hard-core repulsive interactions.
An alternative is the so-called rigid body composite particle approach \cite{nguyen2019aspherical}, which allows for the simulation of arbitrary particle shapes with both attractive and repulsive interactions.
Here, smaller spherical beads are rigidly connected to maintain the overall geometry of a larger, complex particle. 
This strategy offers flexibility, as it can represent a wide range of convex and concave shapes, as well as different surface interactions. 
The interaction between two such particles is commonly computed as a double sum over all pairwise interactions between their constituent beads. 
However, modeling many rigid bodies, each composed of many composite beads, greatly increases computational cost and limits the applicability \cite{nguyen2011rigid} of the method. Additionally, composite bead models can be challenging to construct in a consistent manner due to the large amount of tunable parameters such as composite bead size, spacing, and positioning. 

Significant efforts have been made to reduce the computational cost of calculating attractive and repulsive interactions between rigid body particle pairs. 
For example, the Gay-Berne \cite{berardi1998gay} potential is widely used to simulate ellipsoidal shapes without the need to represent them as composite rigid-bodies. Similarly, 
\citet{ramasubramani2020mean} introduced a Lennard-Jones-like potential applicable to various polyhedral shapes. For both potentials, some limitations on the analytical form and shape of particles exist. In an alternative approach, \citet{lee2020analytical} developed an analytical method for evaluating van der Waals interactions between two faceted particles, avoiding the expensive double-sum and implemented it to accelerate MC simulations \cite{callioglu2025efficient}. 

To reduce computational cost, data-driven approaches have also been explored. 
For example, \citet{fakhraei2025approximation} applied multivariate interpolation to estimate interactions from a limited set of sampled configurations, showcasing the advantages when the model only allows for the generation of small training data sets.
\citet{campos2022machine} developed descriptors for cylindrically symmetric particle pairs, while in Ref.\cite{campos2024machine}, so-called `S functions' were used to construct descriptors for pairs of arbitrary shapes, which can be linearly combined to accurately predict interaction energies. 
\citet{lin2024expanding} extended the Smooth Overlap of Atomic
Positions (SOAP) framework to include ellipsoidal densities in addition to spherical ones. 
\citet{hatch2024anisotropic} simulated large rigid molecules using interpolated interactions from pre-calculated and tabulated data. 

Coarse-graining strategies for anisotropic molecules, such as benzene, include both linear models \cite{nguyen2022systematic} and neural networks \cite{wilson2023anisotropic} optimized to reproduce forces and torques in all-atom MD simulations. 
In our previous work\cite{argun2024molecular}, we employed shape symmetry-aware features fed to a neural network to predict energies, forces, and torques between cube and cylinder particles.
The current challenge is to design an interaction model that captures complex interactions between non-spherical shapes accurately while remaining computationally efficient enough for large-scale simulations.

In this work, we extended the growing machine learning toolkit to model interactions between non-spherical particles. 
Rather than representing each shape by a position and orientation, we described them using a minimal set of points that preserve the particle's symmetry, a more general and flexible approach. 
We then explored machine learning methods designed for regression on point clouds, focusing on Machine Learning Potentials (MLPs) which are typically trained on quantum mechanical data.
We adapted the MLPs commonly used for atomistic point clouds to model interactions between 3D colloidal shapes.
We evaluated the accuracy, inference speed and practicality of several MLPs using a cubic test shape, comparing them with our previous model \cite{argun2024molecular}. 
Among the methods tested, NEP\cite{fan2021neuroevolution} offered the best balance between accuracy and computational efficiency. 
We demonstrated that NEP is applicable to a variety of geometries—including cubes, tetrahedra, pentagonal bipyramids, multi-surfaced cubes, and twisted cylinders—and accurately reproduces the structural properties observed in traditional rigid-body simulations based on explicit distance calculations.

\section{Methods}
\begin{figure}[H]
    \centering
    \includegraphics[width=4cm]{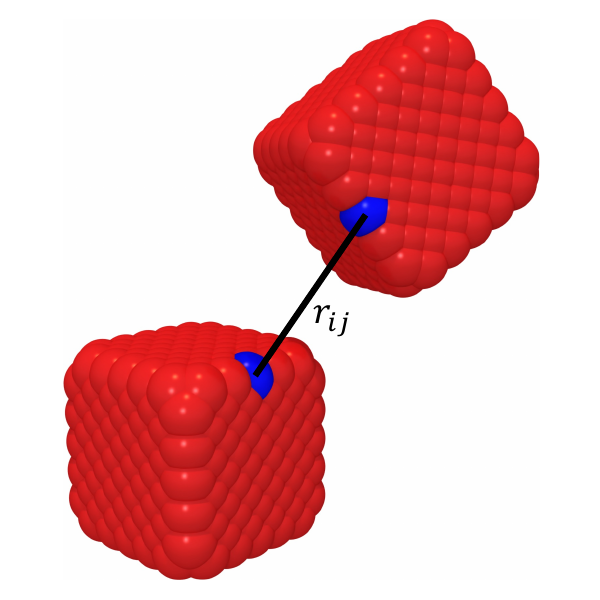}
    \caption{A pair of cubic rigid bodies made of composite beads, with one individual distance $r_{ij}$ resulting in interaction $u_\text{bead}(r_{ij})$ between two composite beads highlighted.}
    \label{fig:pair}
\end{figure}

The overarching objective is to develop a regression model that accurately and efficiently predicts the interaction energy between a pair of non-spherical shapes. 
What distinguishes this task from conventional regression problems is the geometric structure of the input, which originates from 3D Euclidean space. 
In particle-based simulations like MD and MC, systems of rigid bodies are typically described by the center of mass positions ($\mathbf{p}$) and orientation quaternions ($\mathbf{q}$) of the particles.
Accordingly, the basic representation of a rigid body pair, labeled 1 and 2, is given by $\mathbf{x}_{\text{raw}} = (\mathbf{p}_1, \mathbf{p}_2, \mathbf{q}_1, \mathbf{q}_2)$.
The net interaction energy between the two particles is denoted by the function
\begin{equation}
    U = \sum_{i=1}^{N} \sum_{j=1}^{N} u_\text{bead}(r_{ij}) \quad,
\end{equation}
where $u_\text{bead}(r)$ represents the interaction between individual beads as shown in Fig.~\ref{fig:pair}. 
Traditional MD or MC simulations compute this double sum explicitly. 
We aim to approximate the true interaction energy $U$ using a surrogate model $\Phi$, thereby avoiding the computational cost of evaluating the full double sum,
\begin{equation}
    U(\mathbf{x}_{\text{raw}}) = U(\mathbf{p}_1, \mathbf{p}_2, \mathbf{q}_1, \mathbf{q}_2)  \approx \Phi(\mathbf{p}_1, \mathbf{p}_2, \mathbf{q}_1, \mathbf{q}_2) \quad .
\end{equation}
However, $\mathbf{x}_{\text{raw}}$ is not suitable as input for standard regression methods as it does not satisfy the relevant symmetries. 
We categorize these symmetries into two types, as described below, generic (or Euclidean) and shape-specific.

\textbf{Generic (Euclidean) Symmetries}: The target variable, i.e., the output of the true energy function $U$, is invariant under (i) the permutation of the two particles $\mathbf{p}_1 \leftrightarrow \mathbf{p}_2$, Eq.~\eqref{eq:permut_inv}, (ii) a global translation $\mathbf{t}$ applied to both positions, Eq.~\eqref{eq:trans_inv}, and (iii) a common rotation $\mathbf{R} \in \mathrm{SO}(3) $ applied to both particles Eq.~\eqref{eq:rot_inv}: 
\begin{align}
   \Phi((\mathbf{p}_1, \mathbf{q}_1), (\mathbf{p}_2, \mathbf{q}_2)) &= \Phi((\mathbf{p}_2, \mathbf{q}_2), (\mathbf{p}_1, \mathbf{q}_1))\quad, \label{eq:permut_inv}\\
   \Phi((\mathbf{p}_1+\mathbf{t}, \mathbf{q}_1), (\mathbf{p}_2+\mathbf{t}, \mathbf{q}_2)) &= \Phi((\mathbf{p}_1, \mathbf{q}_1), (\mathbf{p}_2, \mathbf{q}_2))\quad, \label{eq:trans_inv}\\
   \Phi(\mathbf{R}(\mathbf{p}_1, \mathbf{q}_1), \mathbf{R}(\mathbf{p}_2, \mathbf{q}_2)) &= \Phi((\mathbf{p}_1, \mathbf{q}_1), (\mathbf{p}_2, \mathbf{q}_2))\quad. \label{eq:rot_inv}
\end{align}
Throughout this paper, we use the terms invariance and symmetry interchangeably.
For a more comprehensive discussion of symmetries in the context of machine learning, we refer the reader to Ref.~\cite{geometric2021}.

\begin{figure}[H]
    \centering
    \begin{subfigure}
        \centering
        \includegraphics[width=8cm]{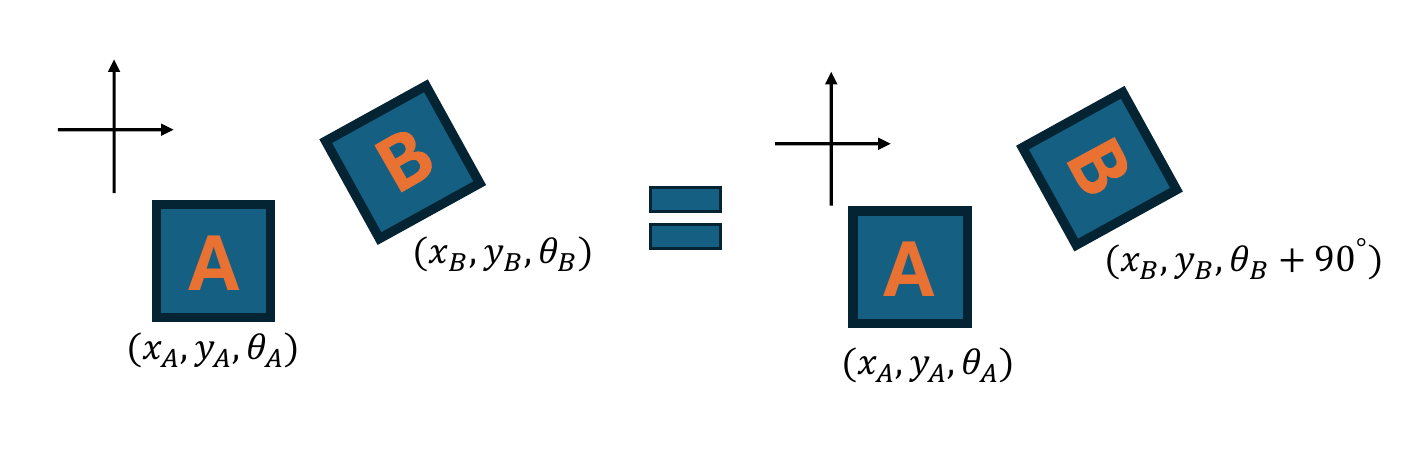}
        \label{fig:shape_sym_b}
    \end{subfigure}
    
    \begin{subfigure}
        \centering
        \includegraphics[width=6cm]{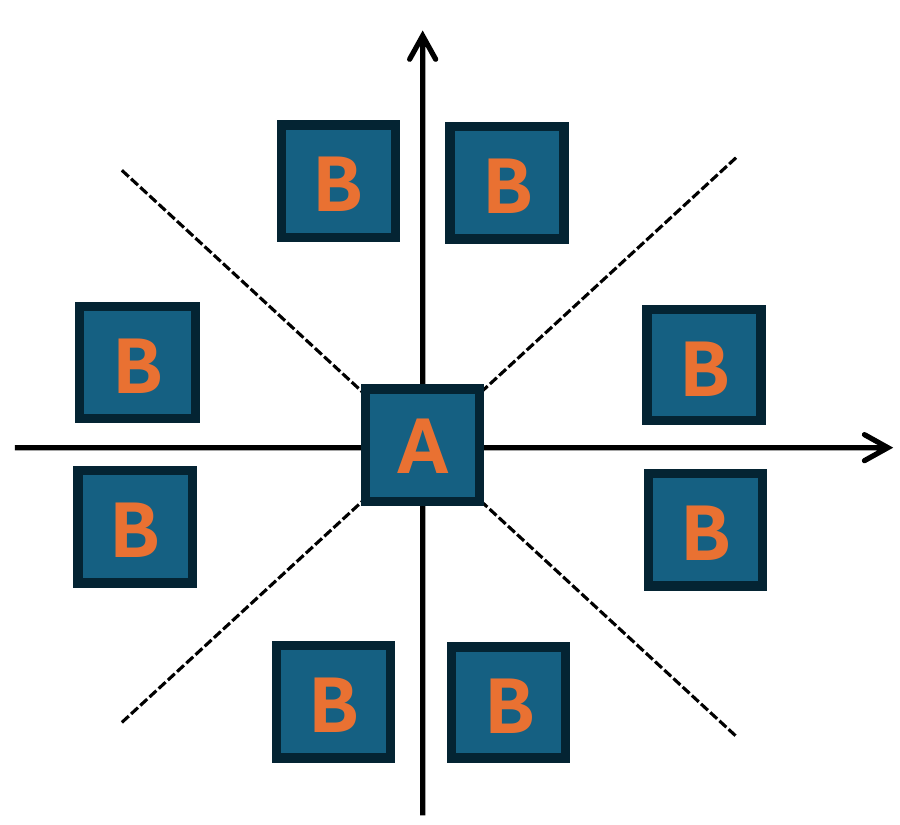}
        \label{fig:shape_sym_b}
    \end{subfigure}
    \caption{Illustration of the symmetries of a pair of squares, where (top) rotating cube \textbf{B}, or (bottom) translating cube \textbf{B} results in the same interaction.}
    \label{fig:shape_sym_both}
\end{figure}

\textbf{Shape Specific Symmetries}: Generic symmetries as described above apply to any pair of particle shapes.
However, for symmetric shapes, i.e., those with point group symmetry, additional continuous (e.g., a cylinder's rotation around its main axis) and/or discrete symmetries must be considered.
Fig.~\ref{fig:shape_sym_both} illustrates the concept of shape symmetries using squares for ease of visualization.
Rotating one square by 90 degrees, as shown in Fig.~\ref{fig:shape_sym_both}(top) or translating cube B into any of the eight possible locations in Fig.~\ref{fig:shape_sym_both}(bottom) results in distinct configurations of $\mathbf{x}_{raw}$ while preserving true interaction energy $U$. 

For a pair of squares, this yields $4\times8=32$ symmetric/invariant states. 
We did not consider improper rotations and the related symmetries since they do not correspond to a real motion in classical mechanics of rigid bodies, in other words, we only considered orientation-preserving symmetries.  
While enumerating the symmetries of two squares is relatively straightforward, the task becomes more complex in three dimensions or with more intricate geometries.
Symmetries of individual geometric objects are very well established in group theory and referred as symmetry groups. 
For example, the orientation-preserving symmetry group of a square, a cube and a tetrahedron are $C_4$, $O$, and $A_4$ respectively. 
However, when we consider the energy of a pair of objects, each with their own degrees of freedom, these symmetry groups cannot be directly used. 

Even after identifying all relevant symmetries, integrating them into a machine learning model $\Phi$ is nontrivial.
Therefore, an efficient approach to account for shape-specific symmetries is highly desirable.

\subsection{Handling Symmetries}
ML approaches that account for symmetries in the data can be broadly classified into three categories: (1) data augmentation, (2) designing symmetry invariant descriptors for the data, (3) using inherently invariant end-to-end models. We describe each approach briefly in the following. 

The first approach involves data augmentation, where symmetric variants of the input data are added to the data set ~\cite{lee2024classification, ronneberger2015u}. 
For instance, considering the square example, each input $\mathbf{x}_{raw}$ in the data set can generate 31 additional configurations by applying the 32-fold symmetry, all sharing the same target energy. 
This method is straightforward if all relevant symmetries can be confidently enumerated. 
However, a key drawback is that the regression model must still learn all these symmetries, as they are not explicitly encoded, making the learning task more complex. 
This approach significantly increases the data set size, leading to higher computational costs during training. 
Furthermore, for continuous symmetries, data augmentation cannot be applied infinitely, which may introduce errors and limit the model's learning capacity.
Given these challenges, we did not consider data augmentation approaches in this work.

The second approach involves engineering symmetry-invariant features by constructing a descriptor $\mathbf{g}(\mathbf{p}_1, \mathbf{p}_2, \mathbf{q}_1, \mathbf{q}_2)$ that explicitly satisfies the generic Eqs.(\ref{eq:permut_inv}--\ref{eq:rot_inv}) as well as shape-specific symmetries. 
Once these invariant descriptors are generated, conventional regression methods can be directly applied.  For example, we have followed this approach in our prior work~\cite{argun2024molecular}; we used `manually crafted` descriptors, where position and orientation of the first body defined a coordinate frame and the position/orientation of the second body was then used to define a pair configuration. More details of this approach are discussed in Section~\ref{sec:invariant-features}.

The third approach leverages inherently invariant models that learn the descriptors dynamically. 
This category primarily includes deep learning architectures and which are also referred to as ``end-to-end'' models, as discussed in more detail in Section~\ref{sec:end-to-end-models}.
In this work, we specifically explored various implementations within the message-passing neural network framework.

Going back to the descriptor-based approach, starting with $\mathbf{x}_{raw}$, a natural initial step is ensuring translational and rotational invariance by computing relative positions and orientations
\begin{align}
    \mathbf{g}_{rel} = (\mathbf{p}_2 - \mathbf{p}_1, \mathbf{q}_2\mathbf{q}_1^*) = (\mathbf{p}_{21}, \mathbf{q}_{21}) \quad . 
\end{align}    
For asymmetric shapes, $\mathbf{g}_{rel}$ is inherently valid \cite{hatch2024anisotropic}. 
Then $\mathbf{p}_{21}$ can be represented in either spherical or Cartesian coordinates, while $\mathbf{q}_{21}$ can be described using a unit quaternion or three Euler angles.

However, practical cases typically involve shapes exhibiting some degree of symmetry.

One approach is to manually account for these symmetries when computing $\mathbf{g}_{rel,shape}$. 
In our earlier study \cite{argun2024molecular}, we adopted this shape symmetry-aware approach, using $\mathbf{g}_{rel,shape}$ as input to a feed-forward neural network. 
Nevertheless, manually identifying and incorporating symmetry information can be tedious, requiring separate analyses for each distinct shape and adding computational overhead that diminishes simulation efficiency.

\begin{figure}[H]
    \centering
    \includegraphics[width=8cm]{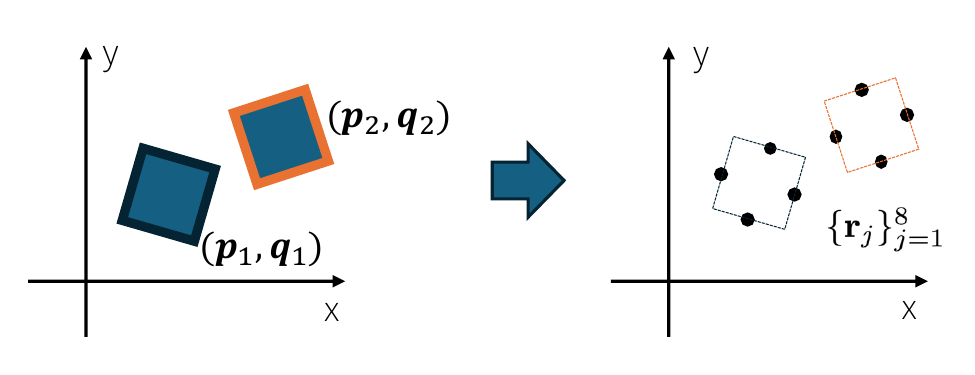}
    \caption{Placing four points at the midpoints of the square's edges preserves the symmetry of the pair of squares. However, point placement is not unique, for example, the corners can also be used. See SI section IIID for further details. }
    \label{fig:shape_to_pts}
\end{figure}

To overcome these challenges, instead of relying on positions and orientations to describe shapes, we propose using a point cloud representation that preserves the same point group symmetry as the shape, as illustrated in Fig. ~\ref{fig:shape_to_pts},

\begin{align}
    \mathbf{g}_{\text{point}}(\mathbf{x}_{\text{raw}}) = \mathbf{g}_{\text{point}}(\mathbf{p}_1, \mathbf{p}_2, \mathbf{q}_1, \mathbf{q}_2) = \{ \mathbf{r}_j \}_{j=1}^{8} 
    \quad . 
\end{align}

The primary advantage of the point-based description is that once the generic symmetries in Eqs.(\ref{eq:permut_inv}--\ref{eq:rot_inv}) are satisfied on $\mathbf{g}_{point}$, the shape symmetries are automatically preserved. 
While a point-based representation is higher-dimensional than a shape-based one, this additional dimensionality serves as the cost of efficiently handling shape symmetries.

Another significant advantage of point-based representations is the ability to leverage existing machine learning methods and tools specifically designed for them. 
Indeed, point clouds are more widely adopted in scientific and engineering applications than the shape-based representation $\mathbf{g}_{rel} = (\mathbf{p}_{21}, \mathbf{q}_{21})$.
In the context of particle simulations, "Machine Learning Potentials" (MLPs) have been employed to construct atomistic potentials and force fields with quantum accuracy. 
MLPs are typically trained using energies computed by density functional theory (DFT) for various atomistic environments or configurations. 
Once trained, these models can infer atomistic interactions without explicitly considering electrons, significantly reducing computational costs.
The input to an MLP is generally the local atomistic environment—specifically, the positions of neighboring atoms, which naturally form a point cloud. 
By leveraging point representations, we can seamlessly apply MLPs to our problem.

\begin{figure*}[t]
    \centering
    \includegraphics[width=\textwidth]{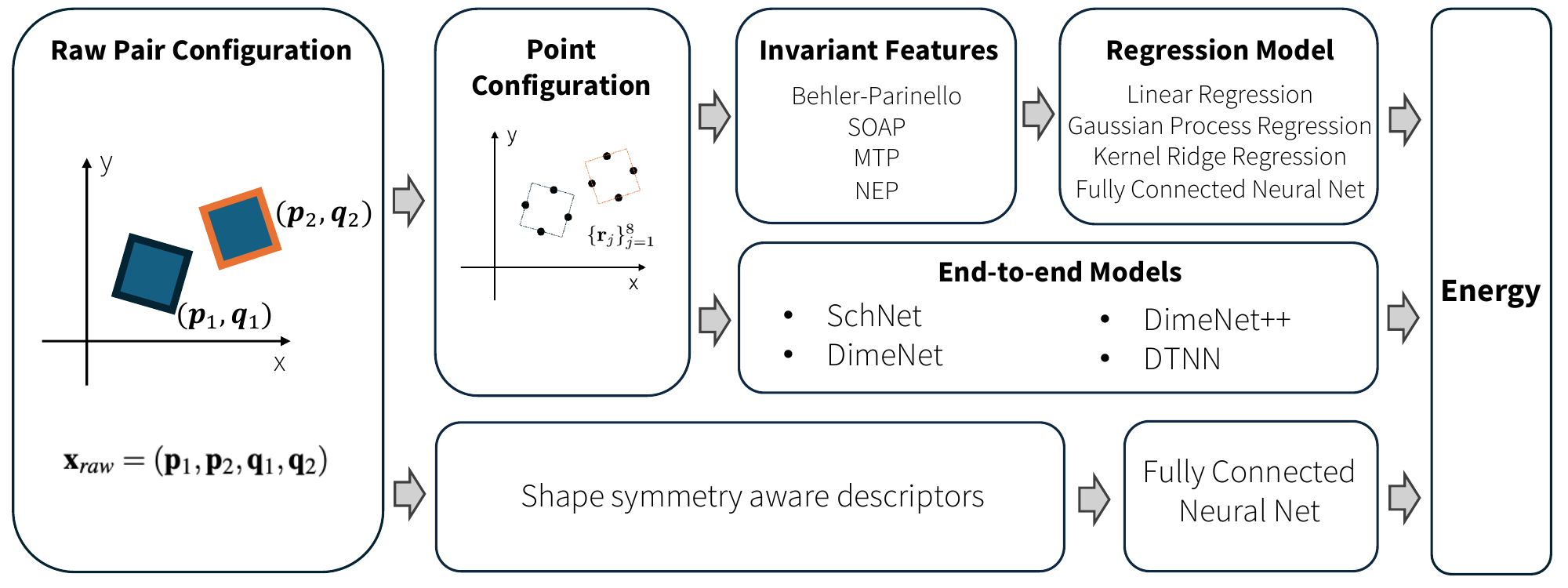}
    \caption{Overview of possible methods to predict energy from raw pair configurations. The raw configuration can be either turned into point clouds, or passed into shape symmetry aware descriptors. Point configurations can then be used with either invariant features (top) or invariant models (middle). The bottom row (shape symmetry aware descriptors + fully connected neural net) corresponds to our prior work \cite{argun2024molecular}.}
    \label{fig:methods}
\end{figure*}
In the remainder of this paper, we evaluated multiple MLP models applied to the point-based representation of a pair of repulsive cubes (for details see SI section I, Fig. S1), comparing their accuracy against our previous shape-based model, which we refer to as the baseline due to its proven capability in reproducing equilibrium structures \cite{argun2024molecular}.
Our comparison focused primarily on two key criteria: accuracy in energy predictions and inference speed. 
Since our goal was to accelerate classical MD or MC simulations of colloidal systems, the deployed model should surpass the efficiency of traditional simulations. 
While many MLP models achieve high prediction accuracy, they often fall short in terms of computational performance. Due to the inherent performance gain of classical MD over most quantum methods, performance is often less of a concern for ML potentials that use quantum training data to build force fields. Here, however, we are directly comparing classical, traditional MD to ML assisted MD.  
Thus, inference speed is a critical factor, particularly as the model needs to be easily parallelizable for GPU acceleration to compete with existing open source MD software \cite{brown2011implementing,pall2020heterogeneous,anderson2020hoomd}.
Secondary considerations include ease of implementation, sensitivity to hyperparameters, training time, and data requirements.
Following the categorization presented in Ref.~\cite{kocer2022neural}, the tested MLPs were grouped into two classes: those with predefined descriptors, i.e descriptor based models, and those with learnable descriptors, i.e end-to-end models. In the following sections we describe, compare, and analyze both approaches in detail. 

\subsection{Descriptor-Based Models }
\label{sec:invariant-features}
A classic example of predefined descriptors are the Behler-Parinello (BP) descriptors, also known as atom-centered symmetry functions \cite{behler2011atom, behler2007generalized}. 
These are constructed using pairwise distances and angles between atomic triplets, which are then fed into a shallow neural network. 
Translational and rotational symmetry is ensured by considering only relative distances and angles, while permutational symmetry is achieved by summing over contributions from neighboring atoms.
Another widely used approach is the Smooth Overlap of Atomic Positions (SOAP) descriptors \cite{bartok2013representing, de2016comparing}, which can also be extended to ellipsoidal particles \cite{lin2024expanding}.
SOAP descriptors can function both as standard feature representations and as kernels, enabling the use of various regression techniques.
In our study, we have explored linear regression, kernel ridge regression, and Gaussian process regression.
Additionally, we have investigated the neuroevolution potential (NEP), which derives its name from the evolutionary algorithm used for training its single-layer neural network \cite{fan2021neuroevolution, fan2022improving, fan2022gpumd}. 
We note that we employed variants of stochastic gradient descent for training in line with common practice.
NEP descriptors share similarities with BP descriptors, utilizing pairwise distances and angles to characterize the local environment.
Specifically, distances are expanded using Chebyshev polynomials, while angles are expanded using Legendre polynomials.
NEP is particularly promising due to its emphasis on computational efficiency (both descriptor generation and inference can be easily parallelized) while maintaining robustness to hyperparameter choices.
In this work, we have implemented the first version, NEP1 \cite{fan2021neuroevolution}. 
Details on data and hyperparameters for all methods can be found in the Supporting Information (SI). 

The accuracy of predefined descriptor-based approaches compared to the baseline method is presented in Table \ref{tab:descriptor_based_accuracy}. We report the mean absolute errors (MAEs) of the energy to quantify accuracy of the predictions. The MAEs are given in units of energy, $\Epsilon$, set by the scale of pair-interaction energies and used later as base unit in MD simulations. 
While BP descriptors fail to achieve adequate accuracy, SOAP descriptors outperform the baseline method when combined with polynomial regression, an extension of linear regression incorporating squared and cubic terms of the SOAP descriptors.
The highest accuracy is achieved using NEP descriptors.

\begin{table}
\caption{Mean Absolute Errors [$\Epsilon$] for the predefined descriptor-based methods on the test data set (repulsive cubes). The baseline method’s error is shown in red. Errors lower than the baseline are shown in bold.}
\begin{tabular}{l||*{6}{c}}{Method}
&\makebox[4em]{MAE [$\Epsilon$]}\\[0.5em]\hline\hline
Baseline Method & \textcolor{red}{0.146} \\\hline
Behler-Parinello + FFNN & 0.240\\\hline
SOAP + Linear Reg. & 0.280\\\hline
SOAP + Polynomial Reg. & \textbf{0.138}\\\hline
SOAP + Kernel Ridge Reg. & 0.280\\\hline
SOAP + Gaussian Process Reg. & 0.262\\\hline
NEP + FFNN & \textbf{0.075} 
\end{tabular}
\label{tab:descriptor_based_accuracy}
\end{table}

Kernel-based methods commonly used with SOAP descriptors are non-parametric, meaning that every prediction requires the entire training data set. 
This significantly limits their inference speed, which made these regression models unsuitable for our purposes. 
However, they were included in this study because they achieve remarkably high accuracy with very small sample sizes, which can be a significant benefit depending on the system of interest. 
As shown in the SI Fig. S4, SOAP with Gaussian process regression achieves the same accuracy with two orders of magnitude less data, while SOAP with polynomial regression  requires one order of magnitude less data, highlighting the strong descriptive power of the SOAP approach.

Another notable MLP is the Moment Tensor Potentials (MTP) ~\cite{shapeev2016moment}. 
Previous comparisons indicated that MTP achieves accuracy comparable to SOAP-based methods while being computationally more efficient ~\cite{zuo2020performance}. 
Its descriptors are powerful enough to be effectively used with linear regression, making it an attractive option. 
However, its descriptor calculations were not easily parallelizable for GPU acceleration, which prohibits the direct use in GPU accelerated MD simulations and therefore not included here.

\subsection{End-to-end Models}
\label{sec:end-to-end-models}
In contrast to the previous section, end-to-end models do not explicitly construct environment descriptors; instead, they implicitly learn them through deep neural networks. 
Most models in this category fall under the Message Passing Neural Network (MPNN) framework, which serves as a general model paradigm for processing graph-like input data rather than a specific deep learning architecture~\cite{gilmer2017neural}.
In the MPNN framework, node/point/atom representations are encoded as hidden embeddings, which interact with neighboring nodes through message and update functions. 
Messages from neighboring nodes are aggregated using a permutation-invariant function, ensuring robustness to the order of node connections.
Through multiple interaction layers, node embeddings are iteratively refined and learned dynamically. 
Typically, after these interaction layers, the hidden representations are passed through fully connected layers to produce scalar outputs. 
These outputs are then aggregated, commonly through summation, maximization, or averaging, to generate the final prediction.
While graphs provide an intuitive representation for (bonded) molecular structures, point clouds can also be transformed into graphs by defining edges between every possible pair of points, enabling similar processing techniques.
Variants of MPNNs arise from different formulations of message and update functions.

We benchmarked three MPNN variants: SchNet~\cite{schutt2018schnet}, DimeNet~\cite{gasteiger2020directional}, and DimeNet++~\cite{gasteiger2020fast}.
SchNet ensures rotational and translational invariance by incorporating pairwise distances between interacting points in its message function. 
DimeNet and DimeNet++ extend this approach by additionally considering angular information from triplets, which further influences the interaction layers.
All models were trained, validated, and tested on the same data set of purely repulsive cubes. 
Details regarding training data set, hyperparameter selection, and implementation specifics are provided in the SI.

\begin{table}
 \caption{Mean Absolute Errors [$\Epsilon$] for the MPNN variants on the test data set (repulsive cubes). The baseline method’s error is shown in red, and the best-performing descriptor-based method (NEP) is shown in green. The lowest error is highlighted in bold.}
\begin{tabular}{l||*{6}{c}}{Method}
&\makebox[8em]{MAE [$\Epsilon$]}\\[0.5em]\hline\hline
Baseline Method$\quad$ & \textcolor{red}{0.146} \\\hline
NEP + FFNN & \textcolor{ForestGreen}{0.075} \\\hline
SchNet & 0.066\\\hline
DimeNet & 0.023\\\hline
DimeNet++ & \textbf{0.015}\\\hline
\end{tabular}
\label{tab:mpnn_accuracy}
\end{table}

In Table \ref{tab:mpnn_accuracy}, we present the accuracy in terms of MAE in energy of various MPNN variants compared to the baseline method.
A fair comparison between MPNNs and predefined descriptor methods is challenging due to the limitations imposed by kernel-based regression methods on training data. 
However, in general, MPNNs outperform predefined descriptor methods in accuracy for the given task, which aligns with previous findings on quantum mechanical data sets \cite{musaelian2023learning}.
When trained with the same amount of data, SchNet achieved half the error of the baseline method, while DimeNet++ improved test errors by an order of magnitude. 
However, inference with MPNNs is more computationally demanding than with BP, NEP, or the baseline methods. 
DimeNet and DimeNet++ were particularly slow due to their complex architectures, which incorporate angular information along with distances. 
This improved accuracy but greatly increased computation time, a key consideration for this application.
SchNet is faster than DimeNet but still slower than the baseline in its current form. 
We used the PyTorch Geometric implementation \cite{FeyLenssen2019} of SchNet with batched GPU inference. 
Despite this, the baseline's neural network inference was at least seven times faster, depending on hardware and batch size. 
An important consideration was also that SchNet avoided the need for explicit descriptor calculations, which are computationally expensive in the baseline method. 
This may, depending on the model size and application, allow SchNet to remain competitive in speed while offering higher accuracy.

After exploring various approaches, we found that descriptor-based methods offered a favorable balance between accuracy and inference speed for the colloidal shapes and interactions examined in this study. 
Although we have shown that end-to-end models tend to be more accurate, their slower inference speed makes them less suitable for simulation use. Because forces and energies must be computed for all pairs of shapes at each timestep during a MD simulation, their significantly lower inference speed in comparison to descriptor-based models is prohibitive for our application. 
However, for more complex shapes and interactions, or in applications where inference speed is less critical, MPNN variants may remain a viable option. 
They can achieve high accuracy with minimal hyperparameter tuning and data pre-processing.
Among the descriptor-based methods, we identified that the NEP+FFNN approach offers the best trade-off between speed and accuracy for applications in colloidal self-assembly. 

\subsection{Molecular Dynamics Validation}

To then rigorously test NEPs performance, we conducted two sets of simulations, one using HOOMD-blue 4.9.1 for conventional molecular dynamics of composite rigid-bodies in an NVT ensemble, and the other using our custom MD code that integrates NEP-predicted forces and torques in NVT. The code is publicly available at \url{https:
//github.com/baho-cb/anisotropic-ML}.

In all simulations, the temperature was kept constant using a Nosé–Hoover thermostat with a standard timestep of 0.005 $\tau$ and a coupling constant of 0.5 $\tau$. 
We used the size of the smaller constituent
beads $\sigma$, as unit of length and mass of a rigid body $m$ as unit of mass. 
Energy is given in units of $\Epsilon$.
$\tau = \sigma\sqrt{m/\Epsilon}$ is the unit of time.
We equilibrated the system for $500 \tau$, before measuring the pair correlation function for $500 \tau$. The particle volume fraction $\phi = N \cdot V_\mathrm{particle}/V$ of all simulations was set to $\phi=0.16$. 

To evaluate the generality of the NEP method, we modeled five distinct particle geometries with repulsive and attractive interactions, unlike the prior performance tests that were completed with purely repulsive cubes for simplicity. 
For shape diversity, we simulated cubes, tetrahedra, and pentagonal bipyramids (shown in the top row of Fig.~\ref{fig:rdf_all}).
To test performance on particles with chemically distinct surfaces, we included a two-bead-type cube (multi-surface cube) with two opposite faces covered in blue beads that interact via purely repulsive forces.
Finally, to assess behavior on completely asymmetric geometries, we modeled a twisted cylinder lacking any point group symmetry (Fig.~\ref{fig:rdf_twisted}).
All simulations employed the same attractive bead–to-bead interaction potential and plotted in the SI Fig. S2. 
The number of constituent beads and exact temperatures of the test simulations are also listed in the SI Table S1. 

\section{Results}

\begin{figure*}[ht!]
    \centering
    \includegraphics[width=\textwidth]{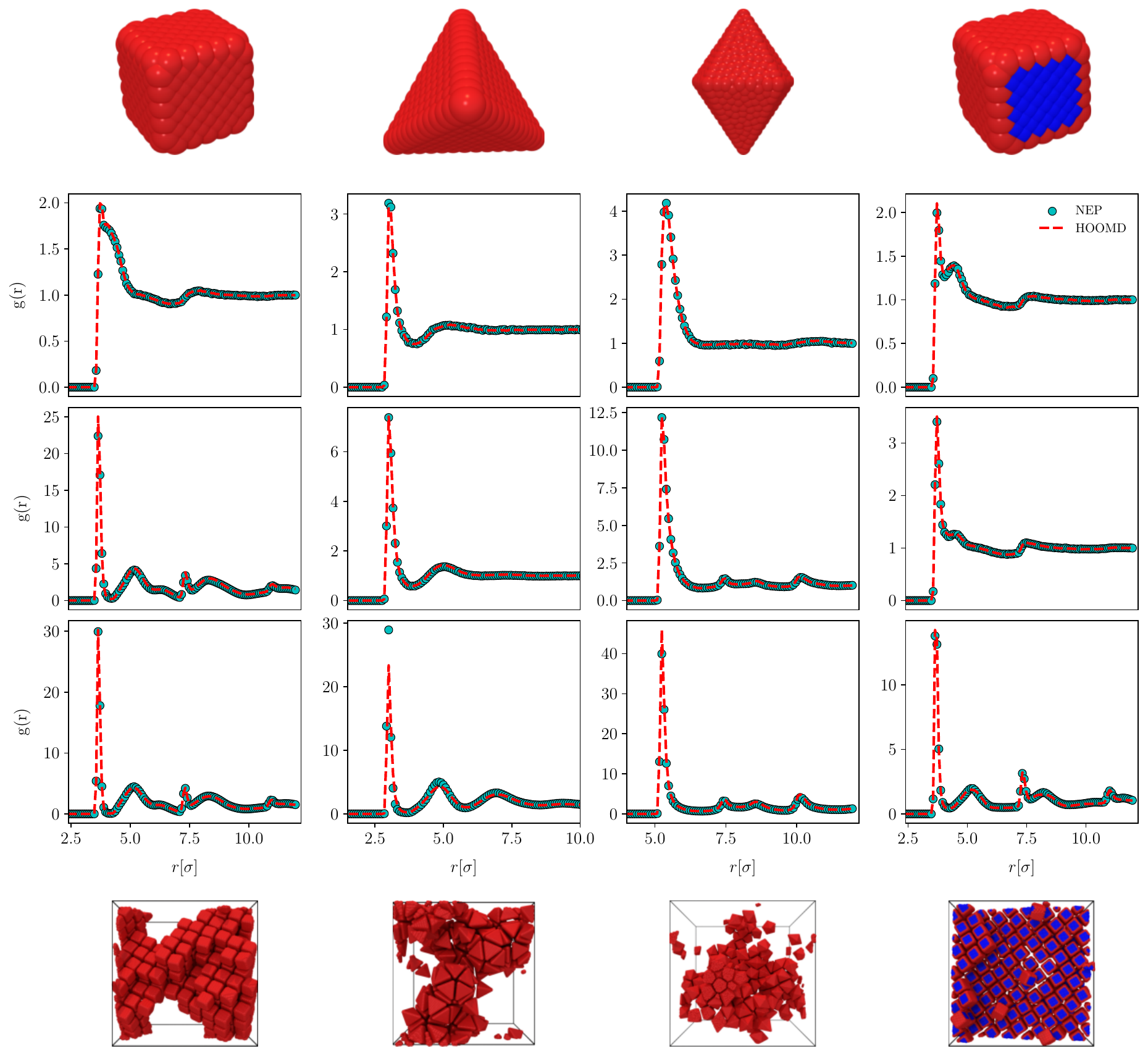}
    \caption{Pair correlation functions $g(r)$ for different shapes are shown in the top, middle, and bottom rows, corresponding to temperatures slightly above, near, and slightly below the transition, respectively. Exact temperatures for each shape are listed in Table S1.}
    \label{fig:rdf_all}
\end{figure*}
To compare NEP and HOOMD simulations, we plotted the pair correlation function of the shapes’ centers of mass at three different temperatures. 
The exact temperatures vary slightly for each shape, chosen to span the transition between dispersed and aggregated states. This range was selected because we expected the pair correlation function to be most sensitive to forces and torques around this transition. 
Specifically, the top, middle, and bottom rows of Fig.~\ref{fig:rdf_all} correspond to temperatures slightly above, near, and slightly below their aggregation transition, respectively. 
Note that we did not aim to determine the nature or exact location of phase transitions in this work.
Across all four shapes, the NEP and HOOMD results showed good agreement in structure, with NEP accurately capturing the structural evolution—from a dispersed suspension at high temperatures to increasingly aggregated and partially ordered configurations at lower temperatures. As expected, the cubes with distinct chemical faces self-assembled into a approximately single layered structure. 

\begin{figure}[H]
    \centering
    \includegraphics[width=8cm]{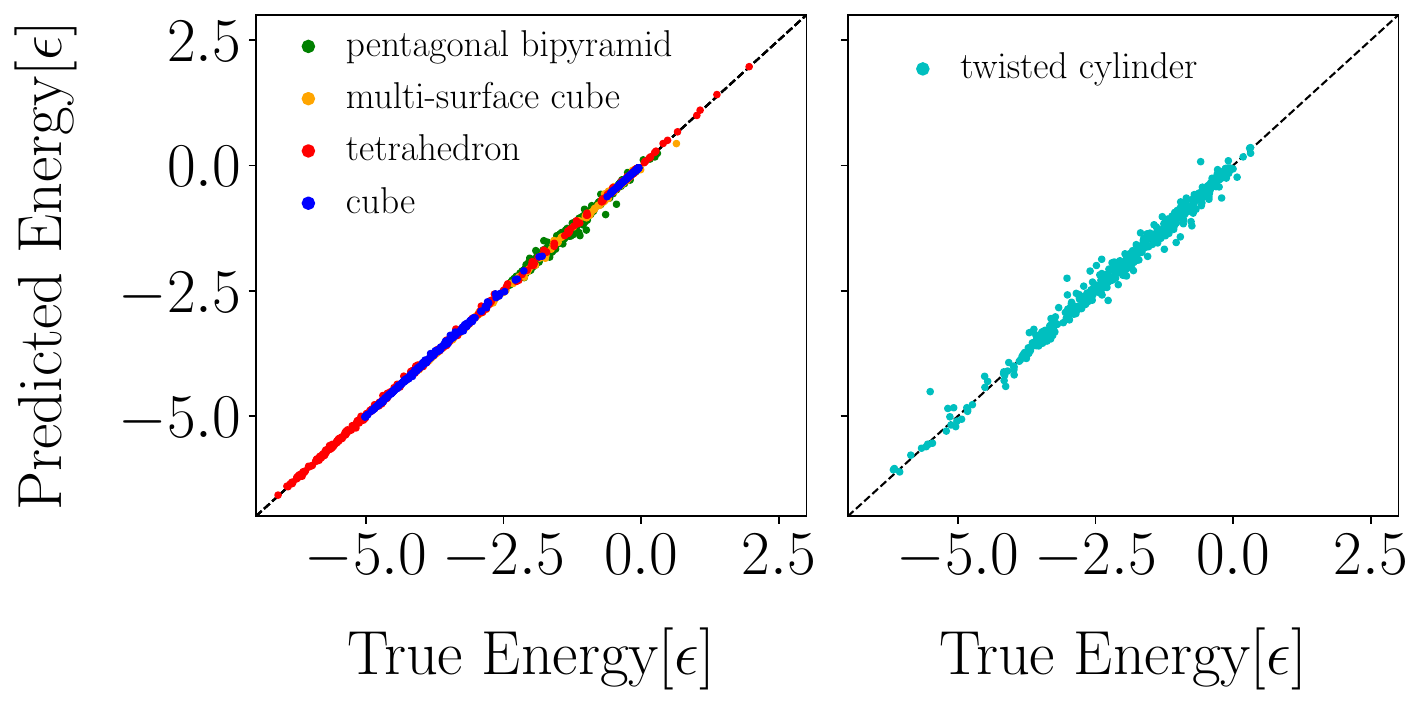}
    \caption{Parity plot of true vs. predicted energy values for the symmetric (left) and asymmetric (twisted cylinder, right) shapes. 10000 test data points are shown for each shape.}
    \label{fig:parity_twisted_vs_others}
\end{figure}

Unlike the four other shapes shown in Fig.~\ref{fig:rdf_all}, the twisted cylinder is completely asymmetric and no rotation, except an identity operation, leaves it unchanged. 
This lack of symmetry increases the complexity of the energy landscape, making the regression task more difficult and resulting in significantly lower prediction accuracy for the twisted cylinder.
In Fig.~\ref{fig:parity_twisted_vs_others}, parity plots for the symmetric shapes are shown on the left while the plot for the twisted cylinder appears on the right.
Prediction errors for each shape are listed in Table~\ref{tab:shape_accuracy}. 
Among the shapes, the cube, tetrahedron, and multi-surface cube exhibited the lowest errors, while the twisted cylinder shows the highest, approximately an order of magnitude larger than that of the cube.
The pentagonal bipyramid fell in between, with error levels closer to those of the symmetric shapes.

\begin{table}[h!]
 \caption{The mean absolute test errors for different shapes using NEP, ordered from most accurate (cube) to least accurate (twisted cylinder).}
\begin{tabular}{l||*{6}{c}}
{Shape} & \makebox[6em]{MAE [$\Epsilon$]}\\[0.5em]\hline\hline
Cube & 0.008 \\\hline
Multi-surface cube & 0.015\\\hline
Tetrahedron & 0.016\\\hline
Pentagonal bipyramid & 0.035\\\hline
Twisted cylinder & 0.079\\
\end{tabular}
\label{tab:shape_accuracy}
\end{table}

\begin{figure}[h!]
    \centering
    \includegraphics[width=8cm]{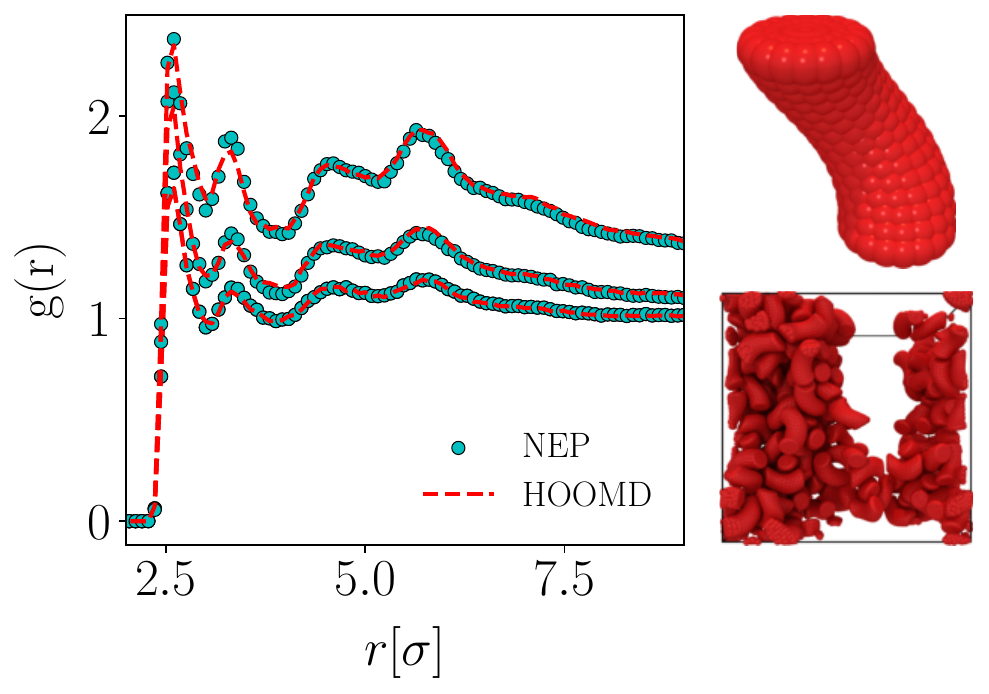}
    \caption{Pair correlation functions of the twisted cylinder shape at three different temperatures, the aggregated structure at the lowest temperature is shown on the right.}
    \label{fig:rdf_twisted}
\end{figure}

Although the error rate for the twisted cylinder are relatively high, the NEP simulations still successfully reproduced its structural features across different temperatures as plotted in Fig.~\ref{fig:rdf_twisted}. 
Unlike the other shapes, the twisted cylinder lacks symmetry and thus did not form ordered aggregates, as indicated by the lower peak intensities in the pair correlation function and confirmed by the configuration at the lowest temperature shown on the right in Fig.~\ref{fig:rdf_twisted}. 
The good agreement in the pair correlation function despite the larger errors in the predicted energies can be partially attributed to this lack of structural order. 

Thus far, forces and torques in the MD simulations were obtained by taking the numerical derivative of the predicted energy from NEP.  
However, it is also possible to calculate the derivatives analytically by applying chain rule as detailed in the SI section 6. Analytical and numerical derivatives led to identical results, as shown in the SI Fig. S7.
While analytical derivation is slightly more challenging to mathematically derive and implement due to many steps involved in the calculation, they are computationally more efficient and do not require parameter adjustment of small displacement and small rotations needed for the numerical derivative approach. 
For performance comparisons that follow, we used the faster implementation with analytical derivatives.

Given our goal was to provide a faster alternative to the composite bead rigid body simulations, we performed a direct performance comparison with the cube shape using analytical derivatives for force and torque calculations. 
While for simulations of smaller size (around 100 particles) there was no significant benefit, NEP outperformed traditional simulations at all larger system sizes (500-4000 particles). 
 NEP can provide almost an order of magnitude speed up depending on the hardware, where we found that for a consumer GPU (RTX 2000) the speedup was between $6\times$ to $9\times$, approaching almost an order of magnitude. On an A100 GPU, speedups ranged from $3\times$ to $8\times$, still a significant performance gain. Additionally, on a single RTX memory severely limits the traditional MD simulation size, whereas NEP assisted MD simulations circumvent this problem. We expect similar limitations to eventually occur for larger systems sizes on the A100.
 In our GPU implementation, the generation of angular descriptors was identified as the bottleneck, while the neural network inference took less than 15\% of simulation time. We point out that we have adopted the original version of NEP~\cite{fan2021neuroevolution} in this work, but there are newer versions\cite{fan2022gpumd} which could improve the computational efficiency of angular descriptor generation for our application. 
We refer the interested reader to Table SI S2 for a detailed performance comparison.

\section{Conclusions}
In this work, we compared several tools to efficiently model the net interaction between two non-spherical particles, represented by rigid bodies of smaller constituent beads. To evaluate different models, we focused on accuracy as well as on performance.
Our overall goal was to provide a feasible alternative to calculating interactions using a costly double sum over all beads. 

We focused on descriptor-based and end-to-end ``machine learning potentials" (MLPs) developed for atomistic systems and quantum mechanical data. We showed that adapting the MLPs to rigid-body pairs was straightforward once particle shapes are replaced with points that have the same point group symmetry as the shapes that they represent.
Our findings highlighted that the descriptor-based methods offer a good balance between accuracy and speed. In particular, the neuroevolution potential (NEP) outperformed other tested combinations.

In order to demonstrate the accuracy and flexibility of NEP when applied to non-spherical colloidal particles, we then modeled various shapes such as cubes, tetrahedra, and pentagonal bipyramids. In addition, we simulated multi-surface cubes and twisted cylinders with no symmetry to showcase the flexibility of our approach. 
For all shapes, NEP MD simulations performed at various temperatures were able to accurately replicate the traditionally obtained pair correlation functions. 
In addition to accurately reproducing structure at various temperatures, we also investigated the performance of NEP MD simulations for cubes. 
Our approach provides a minimum of $3\times$ and up to an order of magnitude speedup with respect to traditional MD simulations, depending on system size and hardware. Additionally, NEP assisted MD simulations enable larger system sizes with less demands on memory.

Although we demonstrated that NEP was both accurate and offered performance improvements at all tested system sizes, several limitations still exist. First, it must be possible to generate a training data set of sufficient size using traditional methods for NEP. However, we also showed that SOAP \cite{bartok2013representing} required overall much smaller training data set sizes for comparable accuracy, which could be a significant benefit depending on the model of interest. Additionally, multivariate interpolation might perform significantly better on limited data sets \cite{fakhraei2025approximation}.
In this study, the model was trained using only energy data. 
However, since analytical expressions for forces and torques have already been derived, they can be incorporated into the cost function. 
This may enhance model accuracy or enable the use of a more compact model, such as fewer descriptors per pair or a smaller neural network, while maintaining the same level of accuracy.

Second, we note that the shapes that we modeled in this study had fairly smooth surfaces, however, our preliminary results suggested that rigid bodies with rougher surfaces and steeper repulsions pose a more challenging regression task. 
The accuracy of the models discussed here depends on the surface roughness and interaction steepness, effects that clearly need further investigations before the methods presented here can be used for more fine grained models such as atomistic systems like rigid proteins, e.g, antibodies, where much steeper interactions and rougher surfaces can be expected.

Further benefits of descriptor-based ML simulations in general could be the additional information that is implicitly synthesized and contained in the descriptors themselves. 
For example, unsupervised learning for structure identification has been successfully applied for simulations of spherical and axial particles \cite{martirossyan2024local, boattini2019unsupervised, adorf2019analysis, lin2024expanding, spellings2018machine}. 
The descriptor-based methods we presented here can similarly be used to classify local environments and to detect various phases or nucleation events in simulations of arbitrary shapes and symmetries. 
Instead of focusing on pairwise descriptors, one could construct descriptors that represent the local environment of anisotropic shapes by including multiple neighbors within a cutoff distance. Exploring this approach is left for future work.

\section*{Acknowledgements}

This work was supported by funding from the Molecule Maker Lab Institute (MMLI): An AI Research Institutes program supported by NSF under award No. 2019897. This work made use of the Illinois Campus Cluster, a computing resource that was operated by the Illinois Campus Cluster Program (ICCP) in conjunction with the National Center for Supercomputing Applications (NCSA) and which was supported by funds from the University of Illinois at Urbana-Champaign.

\section*{Supplemental Information}
The supplementary information provides detailed descriptions of the interactions, data sampling, and preprocessing steps. It also includes implementation and training procedures for the tested MLPs, best practices for applying NEP to non-spherical shapes, mathematical formulations for computing forces and torques, and performance comparison results. The code for running the NEP simulations is available at \url{https://github.com/baho-cb/anisotropic-ML}.

\bibliography{references}

\end{document}


\title{Supplementary Information: Machine-Learning Potentials for Efficient Simulations of Anisotropic Colloids}  

\author{B. Ru\c{s}en Argun}
\affiliation{Mechanical Engineering, 
                              The Grainger College of Engineering, University of Illinois,Urbana-Champaign, 61801, IL}
                
\author{Antonia Statt}
\affiliation{Materials Science and Engineering, 
                              The Grainger College of Engineering, University of Illinois, Urbana-Champaign, 61801, IL}

{
\let\clearpage\relax
\maketitle
}
\tableofcontents

\section{Data Sampling and Pre-processing}
To compare various ML methods, test, training and validation datasets were sampled with latin-hypercube sampling as 6-dimensional pair configurations $\mathbf{g}_{rel} = (\mathbf{p}_{21}, \mathbf{q}_{21})$ where $\mathbf{p}_{21}$ is given in spherical coordinates and $\mathbf{q}_{21}$ is given in Euler angles (`ZXZ' order, intrinsic). 
Once the set of $\mathbf{g}_{rel}$ is obtained, it is straightforward to calculate corresponding true energies and point representations. 

After the initial method comparison, we have created additional dataset to test NEP potential for various shapes. 
These datasets contained around 3 to 4 millions of samples for training and validation depending on the shape.
We noticed that using only latin-hypercube sampling results in under sampling of lower energy configurations. 
Using a model trained only with latin-hypercube samples, the pair correlation functions obtained were not a very good match. 
In addition to latin-hypercube sampling NEP datasets also contains samples from traditional MD simulations to combat the under-sampling of lower energy configurations. 

In the original NEP paper, the cost function includes not only the energy but the force as well.
Since we have derived the analytical expressions for forces and torques for a pair of colloids, in principle it is possible to include them into the training data and cost function. 
This would allow neural net to not only learn the correct energies but also the gradients of the energy and can potentially boost the performance of the NEP approach on colloids. 
\FloatBarrier

\section{Shapes and Interactions}
\begin{figure}[t]
    \centering
    \includegraphics[width=0.5\textwidth]{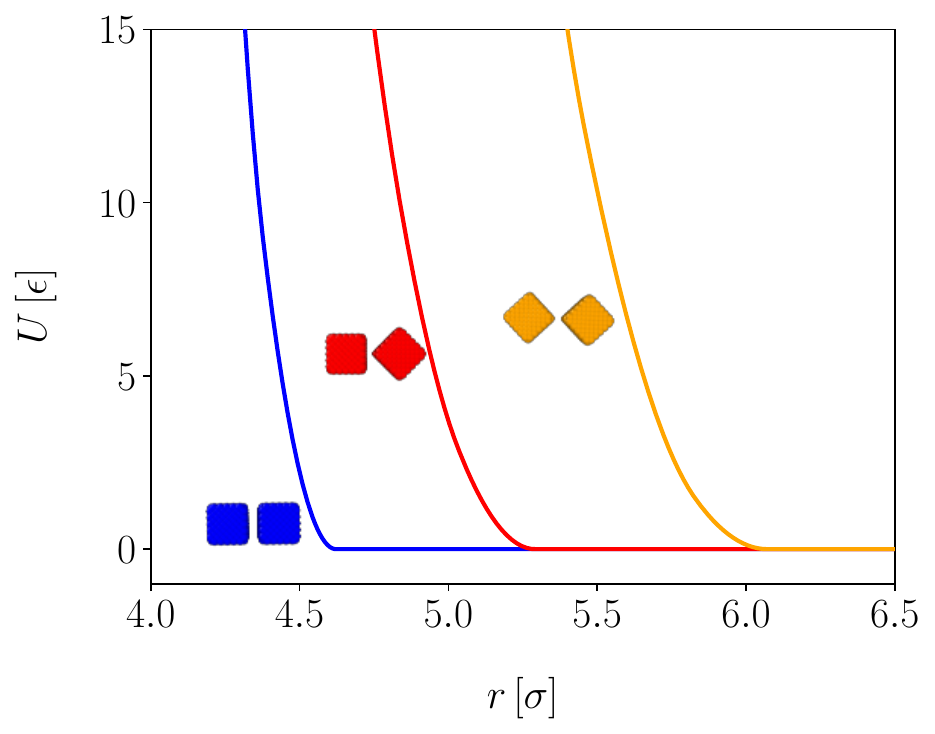}
    \caption{ Repulsive cubes used for initial method comparison. Interaction energy $U(r)$ as function of center of mass distance $r$ for three different cube configurations, as shown next to the curves. Each cube is composed of 150 small spherical beads.}
    \label{fig:wca_cube_interaction}
\end{figure}

\textbf{Purely Repulsive Cube Data Set:}
The test shape used for the method comparison were repulsive cubes composed of 150 beads for ease of computation and fast testing. 
The bead-to-bead interaction was chosen to be purely repulsive, similar to Weeks-Chandler-Andersen (WCA) potential. The resulting total interaction potential between the two cubes at three different relative orientations is plotted in Fig. ~\ref{fig:wca_cube_interaction}.

\textbf{Interactions for MD Validation:} After initial exploration of different methods with the purely repulsive cube data set, we switched to a system that included attractive interactions for a more rigorous test of the NEP simulations. 
Five distinct shapes were modeled as mentioned in the main manuscript. The information on the number of constituent beads they contain, the lowest sampled energy and the simulation temperatures for pair correlations plotted in the main text are listed in Table ~\ref{tab:shape_data}.
\begin{table}[h]
\centering
\caption{Information on systems and temperatures used for both HOOMD-blue and NEP MD simulations.}
\vspace{1em}
\label{tab:shape_data}
\begin{tabular}{l|| c|c|c|c|c}
\textbf{Shape} 
& \shortstack{Number of\\constituent\\beads} 
& \shortstack{Lowest\\energy\\{[$\epsilon$]}} 
& \shortstack{Low temp.\\ {[$\epsilon/k_{B}$]} } 
& \shortstack{Middle temp.\\ {[$\epsilon/k_{B}$]} } 
& \shortstack{High temp.\\ {[$\epsilon/k_{B}$]} } \\
\hline\hline
Cube & 665 & -5.1 & 0.5 & 0.6 & 0.7 \\
Tetrahedron & 455 & -6.7 & 0.5 & 0.6 & 0.7 \\
Pentagonal Bipyramid & 1096 & -2.5 & 0.25 & 0.275 & 0.3 \\
Multi-surface Cube & $500+165$ & -4.4 & 0.4 & 0.45 & 0.5 \\
Twisted Cylinder & 812 & -6.7 & 0.4 & 0.45 & 0.5 \\
\hline
\end{tabular}
\end{table}

\begin{figure}[H]
    \centering
    \includegraphics[width=0.5\textwidth]{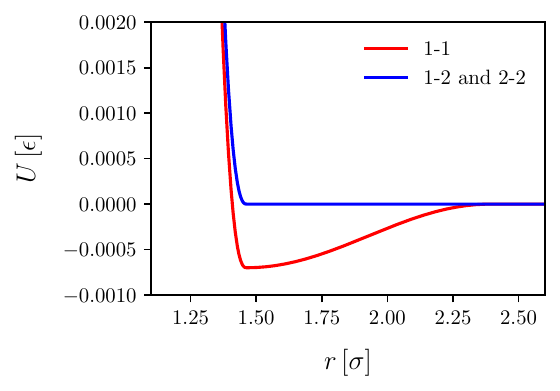}
    \caption{Energy as a function of distance between individual small constituent spherical beads that the rigid body shapes are composed of. All shapes except multi-surface cube contain only one type of beads, denoted by the red $1-1$ line. The multi-surface cube had both interactions.}
    \label{fig:bead_to_bead_attr}
\end{figure}

\begin{figure}[t]
    \centering
    \includegraphics[width=0.5\textwidth]{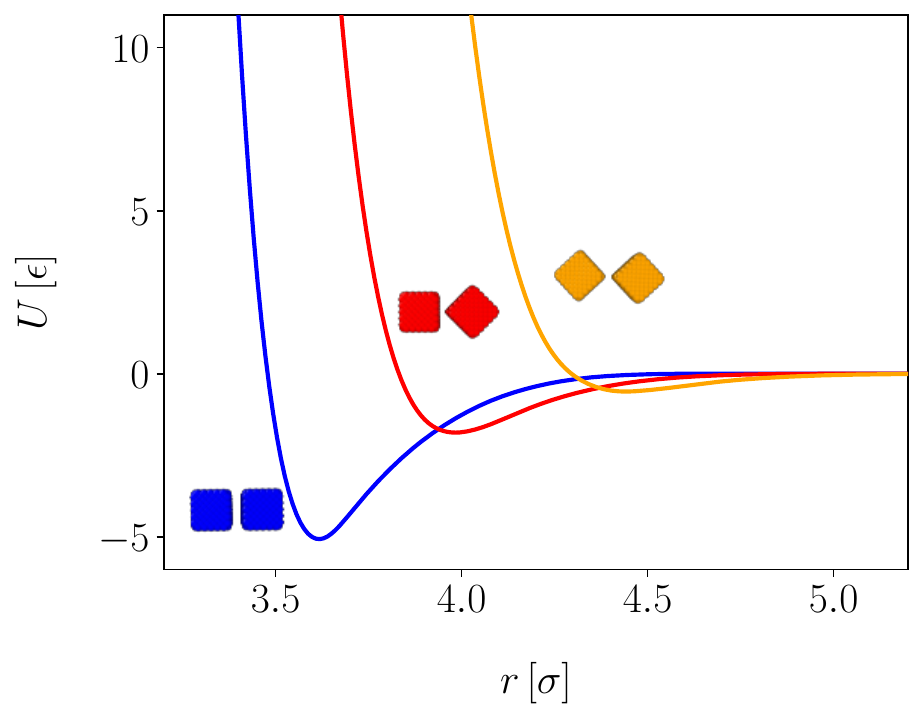}
    \caption{Interaction energy $U(r)$ as function of center of mass distance $r$ for three different attractive cube configurations, as shown next to the curves. Each cube had an approximate side length of 3.2$\sigma$ and are composed of 665 small spherical beads.}
    \label{fig:attr_cube_interaction}
\end{figure}

The constituent bead-to-bead interactions are plotted in Fig. ~\ref{fig:bead_to_bead_attr} and representative resulting overall cube-cube interactions $U(r)$ are shown in Fig.~\ref{fig:attr_cube_interaction}.
The functional form of the potential is 
\begin{equation}
\mathrm{u_{bead}}(r)= \begin{cases}-A \cos (c(r-b)) -A & r>b \\ \epsilon((b-r)/\sigma)^p-2 A & r \leq b\end{cases}\quad,
\end{equation}
with four parameters that somewhat independently adjust the steepness of repulsive part ($p$), depth of the minimum ($A$), position of the minimum ($b$), and the decay rate of the attraction ($c$), i.e., range of the potential. The parameter $\epsilon$ is also used as the unit of energy. 

\FloatBarrier
\section{Predefined Descriptor Methods}
\subsection{Behler-Parinello}
To obtain point based description for a pair of cubes we simply place 6 points to the centers of the 6 faces of each cube. 
We end up with 12 points for each pair of cubes. 
BP symmetry functions describe the environment around a central position i and there are different possible choices for that position. 
We place the central point to the center of mass of the 12 points. 
We have also tried placing the central position to the center of mass of one of the cubes or to any one of the 12 points on the faces but center of mass of the all 12 points gives the best results.  
We use the following symmetry functions from: 

\begin{align}
  &G_i^1 = \sum_j f_c(R_{ij}) \\
  &G_i^2 = \sum_j e^{-\eta (R_{ij} - R_s)^2} \cdot f_c(R_{ij}) \\
  &G_i^4 = 2^{1-\zeta} \sum_{\substack{j,k \neq i \\ \text{all}}} (1 + \lambda \cos\theta_{ijk})^\zeta \cdot e^{-\eta (R_{ij}^2 + R_{ik}^2 + R_{jk}^2)}
\cdot f_c(R_{ij}) \cdot f_c(R_{ik}) \cdot f_c(R_{jk})
\end{align}
where 
\begin{align}
f_c(R_{ij}) =
\begin{cases} 
    0.5 \cdot \left[ \cos \left( \frac{\pi R_{ij}}{R_c} \right) + 1 \right] & \text{for } R_{ij} \leq R_c \\
    0 & \text{for } R_{ij} > R_c
\end{cases}
\end{align}
The index $i$ refers to the central position and the index $j$ runs from 0 to 11 and refers to cube face center positions. 
$R_c$ was set to be equal to the largest distance $R_{ij}$ in the dataset.
We consider the following sets of hyperparameters for candidate $G^2$ generation,
\begin{align*}
&\eta \in \{0.001,0.01,0.1,1.0, 2.0, 4.0, 8.0\}\quad,\\
&R_s \in \{0.0,0.1,0.2,0.3,0.4,0.5,1.0,2.0,3.0,5.0,6.0,8.0,10.0\}\quad. 
\end{align*}
We consider the following sets of hyperparameters for candidate $G^4$ generation, 
\begin{align*}
&\eta \in \{0.001,0.003,0.005,0.007,0.009,0.01,0.015,0.02,0.03,0.05\}\quad,\\ 
&\zeta \in \{1.0,2.0,3.0,4.0,5.0,6.0,8.0\}\quad,\\
&\lambda \in \{1,-1\}\quad.
\end{align*}
Once the candidate descriptors are generated, we need to pick statistically meaningful ones and eliminate redundant sets of descriptors. 
We first eliminate all sets of descriptors that have a standard deviation smaller than 0.05, these descriptors usually give a very close value across most samples so they do not provide useful information to differentiate between different configurations. 
To check for redundancy, we start by adding the first descriptor to the set of final descriptors. 
For every candidate, we calculate a Pearson correlation coefficient with all descriptors already in the final descriptor set.
If the candidate descriptor has a Pearson correlation coefficient higher than 0.95 with any one of the descriptors already in the final set we do not consider it. 
This procedure resulted in 17 radial and 10 angular descriptors, which were then fed in to a fully connected neural network of width 6 and depth 60. 
Increasing the neural net size did not result in any significant improvement on the test accuracy. 

\FloatBarrier

\subsection{Smooth Overlap of Atomistic Positions (SOAP)}

We used the \texttt{dscribe} package v2.1.1 to calculate SOAP descriptors ~\cite{dscribe,dscribe2}. 
Similar to the BP approach, we have added a central point to the center of mass of the 12 points and we have calculated the SOAP descriptors for that central point. 
The cutoff distance in the SOAP calculations was set to be equal to the largest distance from central point to any of the 12 points in the dataset.
The number of radial basis functions and maximum degree of spherical harmonics were both set to 10. 
We found that increasing these further did not result in significant improvement in accuracy. 
Finally, the ideal width for the Gaussian was found to be in the 0.3-0.7$\sigma$ range. 
We note that for both methods it is generally possible to improve accuracy by selecting different point types for the points belonging to different rigid bodies, and leave this direction for further work. 

\begin{figure}[H]
    \centering
    \includegraphics[width=0.6\textwidth]{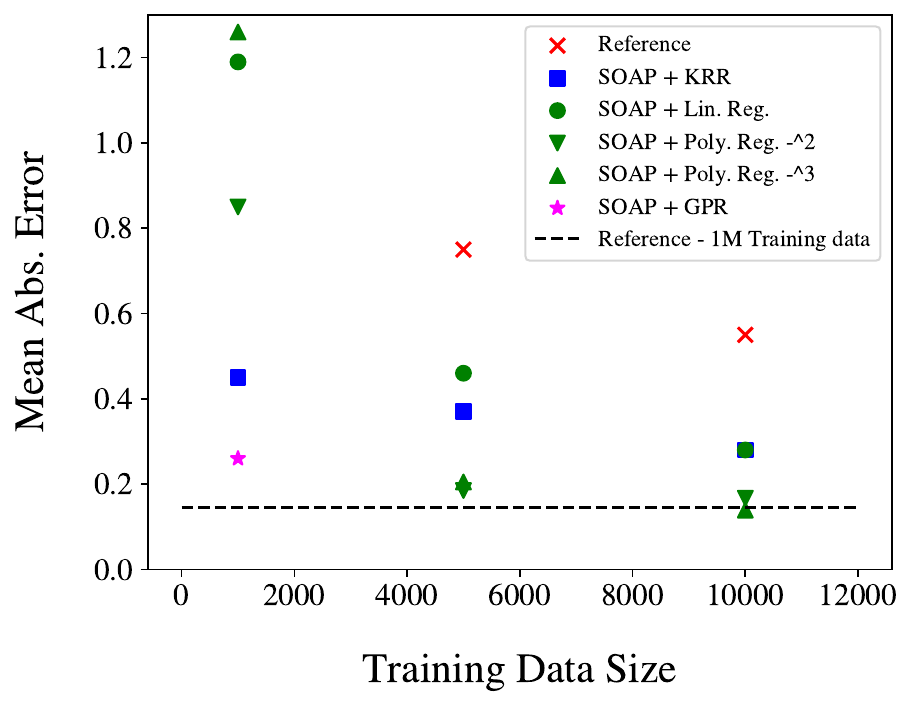}
    \caption{Accuracy of various regression methods with SOAP descriptors as a function of training samples used. The dashed black line refers to the reference method from~\cite{argun2024molecular} with a 1M data set, and red crosses correspond to the same method with data sets of 5000 and 1000 size.}
    \label{fig:soap_vs_trainsize}
\end{figure}

Various regression methods were tested with SOAP descriptors including kernel-based ones, even though in practice they are slow in inference step and limited in training data size. 
The results are plotted in Fig.~\ref{fig:soap_vs_trainsize}.
For kernel ridge regression (KRR), optimal regularization was found to be $\alpha=0.001$. 
For Gaussian process regression\cite{rasmussen2014gaussian} (GPR), we obtained the best results using a combination of constant kernel (with constant 1.0) and radial basis function kernels (with length scale 1.0).

\subsection{Neuroevolution Potential (NEP)}
Similar to the other descriptors we have first placed the central point at the center of mass of the 12 points.
Following from ~\cite{fan2021neuroevolution}, radial descriptors were calculated as, 
\begin{align}
G_{rad,n}^i &= \sum_{j \neq i} g_n(r_{ij}) \label{eq:g_rad}\\
g_n(r_{ij}) &= \frac{T_n(x) + 1}{2} f_c(r_{ij}) \label{eq:chebysev}\\
x &= 2\left(\frac{r_{ij}}{r_c} - 1\right)^2 - 1 
\end{align}
with
\begin{align}
f_c(r_{ij})=
\begin{cases}
    \frac{1}{2} \left( 1 + \cos \left( \pi \frac{r_{ij}}{r_c} \right) \right) & \text{for } r_{ij} \leq r_c \\
    0 & \text{for } r_{ij} > r_c
\end{cases}
\end{align}
The index $i$ refers to the central position and the index $j$ runs from 0 to 11 and refers to cube face center positions. 
$T_n(x)$ is the first-kind Chebysev polynomial of degree $n$.  
$r_c$ was set to be equal to the largest distance $r_{ij}$ in the dataset. 
For angular descriptors the following equation was used, 
\begin{align}
G_{nl}^i = \sum_{j \neq i} \sum_{k \neq i} g_n(r_{ij}) g_n(r_{ik}) P_l(\cos\theta_{ijk}) \label{eq:g_ang}
\end{align}
The double sum runs over every possible triplet of $ijk$ where i is the central position and $jk$ are cube face points. 
$P_l$ is Legendre polynomial of degree $l$. 
After setting $r_c$ 3 remaining hyperparameters are $n_{max}^R, n_{max}^A, l_{max}$. 
These hyperparameters define the number of descriptors by setting an upper bound to the degree of polynomials to be calculated, thus higher values will result in a higher number of descriptors per pair configuration and a higher accuracy. 
However, descriptor calculation is also the performance bottleneck during the actual simulations, so depending on the system simulated and the accuracy desired in the simulations, these numbers must be tuned to find the right balance between the speed and accuracy. 
Chebysev polynomials are calculated starting from degree 0 up to $n_{max}^R$ and $n_{max}^A$ for radial and angular descriptors, respectively, while Legendre polynomials start from degree 1 up to $l_{max}$. Thus the total number of descriptors $n_{\text{desc}}$ is, 
\begin{align}
n_{\text{desc}} = (n_{\max}^{R} + 1) + (n_{\max}^{A} + 1) l_{\max}
\end{align}
We find the set of $n_{max}^R, n_{max}^A, l_{max}$ that can accurately reproduce the equilibrium pair correlation functions across the transition temperature range to be $10,8,4$ which gives 47 descriptors per pair configuration. 
For the cube however $8, 8, 4$ which gives 29 descriptors is powerful enough to get a good match in pair correlation functions.  
These numbers will change depending on the specific interactions, shape complexity and desired accuracy. 
We do not rescale or normalize the descriptors before feeding to the neural network. 
Target variable (energy) is scaled to $[0,1]$ range. 
We train a fully connected neural network of width 150 and depth 3, ReLU is used as the activation function. 
We use a batch size of 1024 and a learning rate of 0.0001 with Adam optimizer. 
We train the model for 500 epochs and pick the model with the lowest mean absolute error on the validation set to perform simulations with. 

\FloatBarrier
\subsection{Point Placement}
The comparison between different methods and approaches was done with the purely repulsive cubes. 
The minimal number of distinct points that have octahedral symmetry is six, and we place them on the cube onto the positions $\left\{ (\pm a,0,0), (0, \pm a,0), (0,0, \pm a) \right\}$, where $(0,0,0)$ is the origin of the cube.
An intuitive choice for $a$ would be half of the cube's edge length, which places the points directly on the faces of the cube.  However, after preliminary studies on different shapes, we believe that $a$ should be treated as a hyperparameter which needs be tuned as well. We found that a range from 
$0.8r_\text{shape}$ to $1.5r_\text{shape}$ was a good starting point to tune $a$, where $r_\text{shape}$ is the characteristic radius of the shape of interest. The range was tuned as part of hyperparameter optimization for each shape. 

Another valid point placement would be placing eight points to the eight vertices of the cube, since full octahedral symmetry would be still ensured. However, using higher number of points increased the computational costs in the descriptor generation stage. 
Therefore, the optimal placement is the one with the minimum number of points that fully covers all symmetries of the shape to be modeled.

\FloatBarrier
\section{Multi-surface Shapes}
To demonstrate the capabilities of the NEP approach we have simulated a cube shape where 2 opposite faces are covered with beads that interact with purely repulsive interaction in addition to attractive beads. 
As expected, these multi-surface cubes aggregate into flat sheets and form a square lattice structure with non-attractive faces pointing out of the plane (Fig. 5 last column).
NEP method is designed for atomic systems, so they are capable of differentiating between different elements when describing the local environment. 
A simple way to take the element type into account is to weight the individual terms (eq.~\ref{eq:chebysev}) of the descriptor functions by the element number of the atoms $i$ and $j$ before summing them up (eqs.~\ref{eq:g_rad},~\ref{eq:g_ang}). 
In our case, the type of the central position $i$ is irrelevant since it is always the center of mass of the pair. 
For shapes with same surfaces all shape positions $j$ are equivalent so no modification to the NEP descriptors is needed. 
The multi-surface cube is expressed with 6 points like the regular cube, however the points on the non-attractive faces assumed to have a different type and their contributions to the descriptors, i.e. $g_{n}(r_{ij})$ where $j$ is a non-interacting surface point, is weighted by 2. 

To simulate systems with more than one shape, a similar approach can be utilized.
Points representing different shapes can be weighted differently. 
For example, for a pair of a cube and a tetrahedron, the points of the cube can be type 1 and the points of the cube can be type 2.
Thus $g_{n}(r_{ij}) , j \in cube$ will be multiplied by 1 and  $g_{n}(r_{ij}) , j \in tetradedron$ will be multiplied by 2. 
Whether the same neural network can be used to model shape 1 - shape 1, shape 2 - shape 2 and shape 1 - shape 2 needs to be tested. 

\FloatBarrier
\section{Asymmetric Shapes}
Pairs of shapes with no point group symmetry pose a more challenging regression task. 
We use the twisted cylinder as an example. 
An intuitive choice for point placement for an arbitrary shape is to put 3 linearly independent points, for example $\mathbf{r} = \left\{ (1,0,0), (0,1,0), (0,0,1) \right\}$ where origin is the center of mass of the shape.
This point placement resulted in significantly higher error rates compared to other shapes. 
We have tried weighting the terms of distinct points differently, but the results did not improve by much. 
For all shapes considered in this paper the initial orientation is picked such that the principal axis of the shape are aligned with the Cartesian directions (i.e. the moment of inertia tensor is diagonalized). 
Scaling the point positions $\mathbf{r}$ by the inverse of diagonal terms of the moment of inertia slightly improved the accuracy. 
For the twisted cylinder best results were obtained by placing points along the twisted axis of the cylinder.
Our goal with this shape is to demonstrate how the methods extends to arbitrary shapes and we recognize that this specific point placement is exploiting a regularity of the shape even though it is not a proper point group symmetry.  
The main outcome is that placing points to fill in the shape; which is doable for arbitrary shapes, just not as straightforward as it is the case for twisted cylinder whose axis can be given analytically, can provide adequate accuracy. 

\FloatBarrier
\section{Force and Torque calculations}
\begin{figure}[H]
    \centering
    \includegraphics[width=0.7\textwidth]{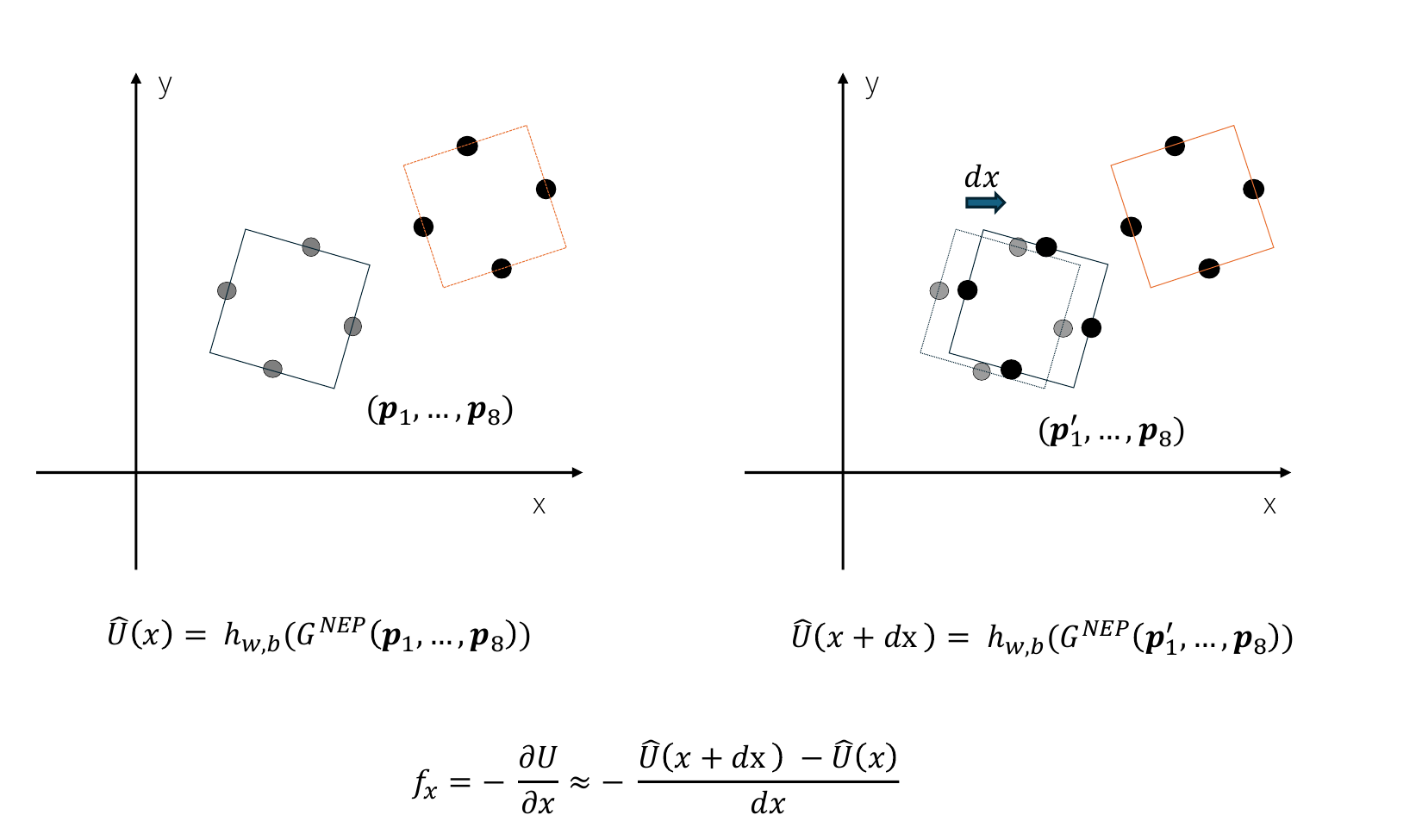}
    \caption{Illustration of the procedure for numerical force calculation.}
    \label{fig:si_force}
\end{figure}

\begin{figure}[H]
    \centering
    \includegraphics[width=0.7\textwidth]{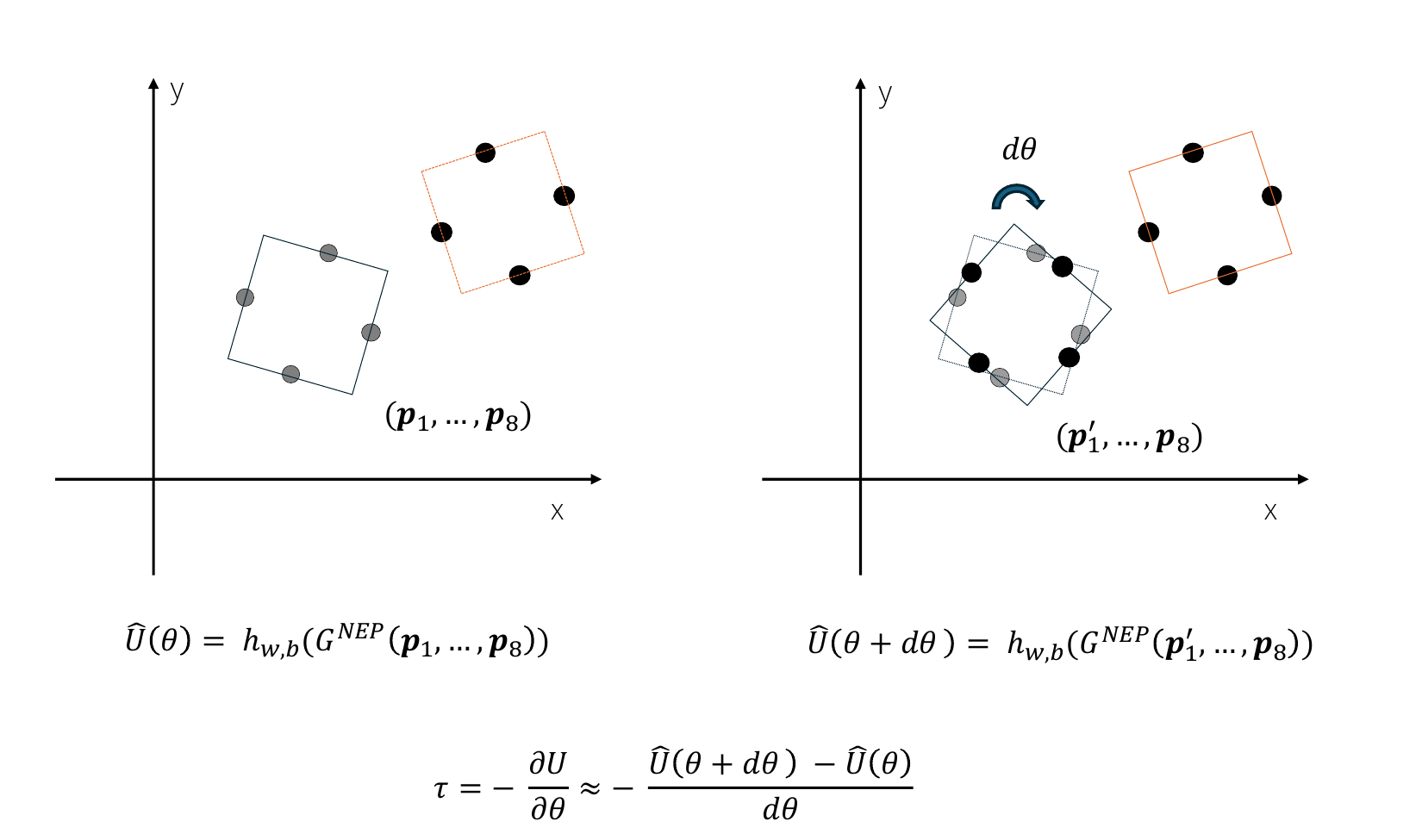}
    \caption{Illustration of the procedure for numerical torque calculation.}
    \label{fig:si_torque}
\end{figure}

All models considered in this study are invariant models meaning they cannot be directly used to predict forces and torques which are equivariant quantities. 
To illustrate, when the pair configuration is rotated the energy stays the same but the resulting force vector is rotated by the same amount. 
Torque and force vectors can be calculated by taking the derivatives of the energy with respect to the pair configurations of the rigid bodies. 
The derivatives can be calculated numerically or analytically. 
Numerical calculations are simpler to implement as illustrated in Figs. 
\ref{fig:si_force} and \ref{fig:si_torque} in 2D. 
The accuracy of approximated forces and torques depend on the $dx$ and $d\theta$. 
We find the optimal range is between $0.001$ and $0.00001$. 

\begin{figure}[H]
    \centering
    \includegraphics[width=\textwidth]{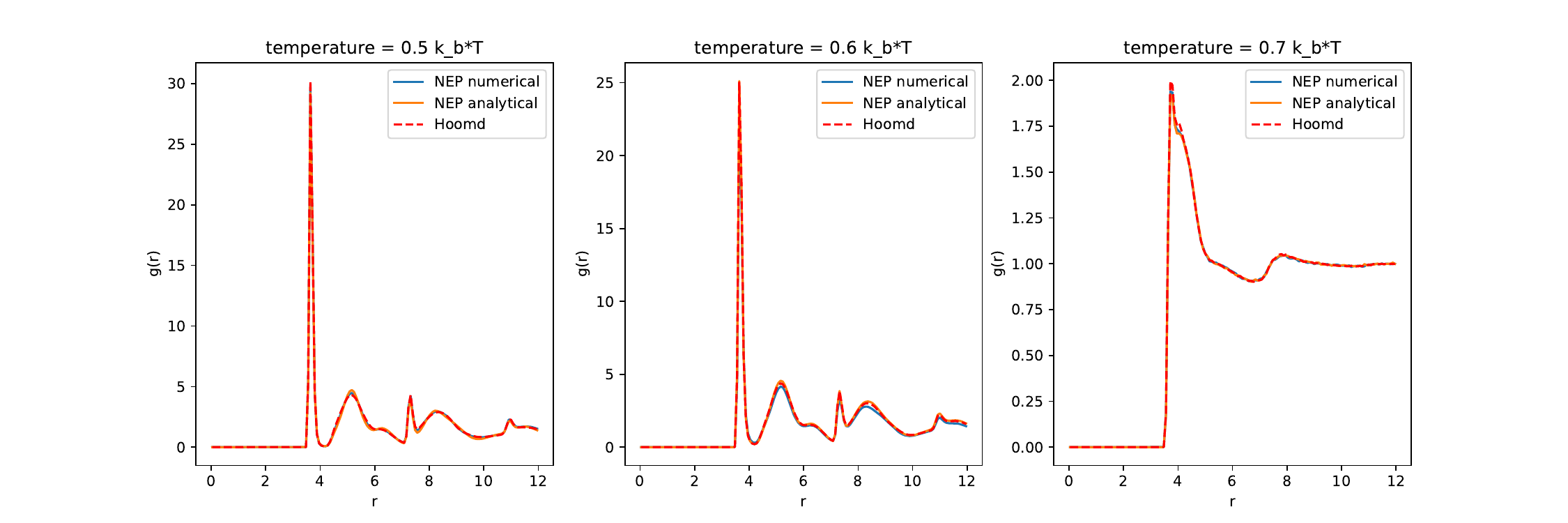}
    \caption{Pair correlation function at three different temperatures for cubes. Both NEP simulations, with analytically and numerically calculated forces-torques, match  HOOMD simulations. 29 NEP descriptors (8, 4, 4) were used for both numerical and analytical NEP.}
    \label{fig:analytical_gr}
\end{figure}

Calculating analytical derivatives, which is effectively applying the chain rule to the energy prediction, requires more implementation effort but is computationally more efficient and eliminates the need for additional parameters such as $dx$ and $d\theta$. 
The corresponding analytical expressions can be found in the shared simulation code \url{https://github.com/baho-cb/anisotropic-ML}.
Pair correlation functions for simulations using numerical derivatives are shown in Fig. 5 of the main text. 
To validate the analytical derivative approach, we performed simulations with cubes and observed good agreement in the pair correlation functions when compared to HOOMD results as shown in Fig. ~\ref{fig:analytical_gr}.

\FloatBarrier
\section{Performance Comparison}

Simulation performance measured in timesteps per second (TPS) primarily depends on hardware and system size, $N_\text{particles}$. 
For NEP simulations, the number of descriptors $n_\text{desc}$ also affects evaluation speed. 
Higher $n_\text{desc}$ values generally yield better accuracy but lower performance.

We benchmarked NEP and traditional HOOMD simulations on two GPUs, A100 and RTX 2070 Super, across four system sizes: $N_\text{particles}=500$, 1000, 2000, and 4000. 
For NEP, we tested two descriptor settings: $n_\text{desc} = 47$ (with $n^{R}_{\max}=11$, $n^{A}_{\max}=4$, $l_{\max}=4$) and $n_\text{desc} = 29$ (with $n^{R}_{\max}=8$, $n^{A}_{\max}=4$, $l_{\max}=4$). 
The latter was used in the cube simulations shown in Fig. 5 of the main text.

To ensure a fair comparison, both NEP and HOOMD simulations were run in single precision. 
TPS results are reported in Table S2. 
NEP consistently outperformed HOOMD, achieving speedups of several fold, with a maximum improvement approaching an order of magnitude. 
This performance gain stems in part from NEP's ability to bypass the need to track the positions of composite beads, reducing memory usage and enabling simulations of larger systems which is an important advantage on memory-constrained GPUs like the RTX 2070.

\begin{table}[h!]
 \caption{Timesteps per second of NEP simulations compared to HOOMD-blue for various system sizes $N_{particles}$. The approximate speedup achieved is also noted. A dash (---) indicates that GPU memory was not sufficient to run the simulation.}
\begin{tabular}{l||*{6}{c}}
\backslashbox{GPU - Method}{$N_{particles}$}
&\makebox[6em]{500}&\makebox[6em]{1000}&\makebox[6em]{2000}&\makebox[6em]{4000}\\[0.5em]\hline\hline
RTX-2070 - HOOMD & 35 & 18 & --- & ---\\\hline
RTX-2070 - NEP ($n_{desc} = 29$) & 290 & 166 & 86 & 44\\\hline
RTX-2070 - NEP ($n_{desc} = 47$) & 212 & 116 & 57 & --- \\\hline
Approx. Speedup & $6\times$ to $8\times$ & $6\times$ to 9$\times$ & -- & --\\[0.5em]\hline
A100 - HOOMD & 105 & 61 & 29 & 16.2 \\\hline
A100 - NEP ($n_{desc} = 29$) & 375 & 375 & 237 & 134 \\\hline
A100 - NEP ($n_{desc} = 47$) & 380 &  300 & 173 & 95 \\\hline 
Approx. Speedup & $3\times$ to $4\times$ & $ 5\times$ to $6\times$ & $6\times$ to $8\times$ & $6\times$ to $8\times$\\[0.5em]\hline
\end{tabular}
\label{tab:performance}
\end{table}

We note that in our previous work ~\cite{argun2024molecular}, TPS measured for HOOMD simulations were different since HOOMD version 2.9.4 was used with mixed precision while in this work we use version 4.9.1 with single precision.  

\FloatBarrier

%